\documentclass[12pt]{article}
\usepackage[margin=1in]{geometry}                 
\usepackage{setspace}
\usepackage{graphicx}
\usepackage{amsmath}
\usepackage{amssymb}
\usepackage{bm}
\usepackage{url}
\usepackage{epstopdf}
\usepackage{times}
\usepackage{courier}
\usepackage{authblk}

\usepackage{algorithmic}
\usepackage{textcomp}
\usepackage{xcolor}
\usepackage{subcaption}
\usepackage{diagbox}
\usepackage{tabulary}
\usepackage{mathtools}
\usepackage{listings, multicol}
\definecolor{codegreen}{rgb}{0,0.6,0}
\definecolor{codegray}{rgb}{0.5,0.5,0.5}
\definecolor{codepurple}{rgb}{0.58,0,0.82}
\definecolor{backcolour}{rgb}{0.95,0.95,0.92}

\lstdefinestyle{mystyle}{
    backgroundcolor=\color{backcolour},   
    commentstyle=\color{codegreen},
    keywordstyle=\color{magenta},
    numberstyle=\tiny\color{codegray},
    stringstyle=\color{codepurple},
    basicstyle=\ttfamily\footnotesize,
    breakatwhitespace=false,         
    breaklines=true,                 
    captionpos=b,                    
    keepspaces=true,                 
    showspaces=false,                
    showstringspaces=false,
    showtabs=false,                  
    tabsize=2
}

\title{Randomness Is All You Need: Semantic Traversal of Problem-Solution Spaces with Large Language Models}
\author{Thomas Sandholm, Sayandev Mukherjee, Bernardo A. Huberman}
\affil{NextGen Systems, CableLabs, Santa Clara, CA}

\begin{document}
\maketitle

\begin{abstract}
We present a novel approach to exploring innovation
problem and solution domains using LLM fine-tuning
with a custom idea database.
By semantically traversing the bi-directional
problem and solution tree at different temperature
levels we achieve high diversity in solution edit distance
while still remaining close to the
original problem statement semantically.
In addition to finding a variety of solutions to a given
problem, this method can also be used to refine and clarify the original
problem statement. As further validation of the approach,
we implemented a proof-of-concept
Slack bot to serve as an innovation assistant.
\end{abstract}

\section{Introduction}\label{sec:introduction}
Innovation is a creative process that while hard to replicate, it involves structure, processes, and workflows that allow innovators to learn from past experiences to simplify creation and validation of new ideas. Too many processes and
structures may however reduce the novelty and impact of new ideas and thus lead to slow innovation progress.
As an example, if innovators find the process of filling in the fields of a Lean Canvas\footnote{\url{https://www.leanfoundry.com/tools/lean-canvas}} questionnaire to be too daunting, the innovation pipeline of new ideas may dry up.  

It is in the nature of good innovation ideas that they may have been suggested before in
similar forms, possibly in related fields.  However, innovators seldom perform due diligence to catch closely-related or similar ideas, either due to time constraints or to avoid getting tainted by exposure to ``prior art'' from a patent perspective.
We believe there is a latent mechanism that seasoned innovators in a particular field use
to suggest solution approaches worth exploring given a particular problem. Given there often exists a large database of prior innovations in an institution, one would like to access them in a format that can assist innovators working on given problems. While we do not expect to fully automate innovation, we propose an AI-tutor or collaborator to guide
an innovator through the ideation process. This not only involves discovering prior art but actually suggesting
novel solution approaches. Previous work has found that such unbiased bots have a positive effect on creativity~\cite{hwang2021}.

Large Language Models (LLMs) are good at parsing through and making sense of large bodies of text and producing creative output. Moreover, the democratization of foundation models and efficient
fine-tuning techniques make it possible to take published pre-trained models (that were trained on public datasets)
and fine-tune them locally on proprietary data containing private or sensitive information that should be
kept out of the public domain.

In this paper, we consider a scenario involving internal innovation proposals in an enterprise, such as an idea submissions
database.  While innovators do have access to all previous innovations, their searches tend to be restricted to traditional keyword or tag matches. Worse, even after finding relevant projects it is time consuming to 
read through all projects and extrapolate the lessons learned.

The thought process involved in problem solving is non-linear and tends to branch out into alternative exploration paths with
the ability to backtrack. Mimicking this process has been shown to improve LLM reasoning capabilities~\cite{wei2022}.

We adopt a similar chain-of-thought, tree-like structure where problems are mapped to solutions and solutions to problems
via local fine-tuned LLMs. This tree model is further enhanced with branches of semantically similar problems
from an internal idea database. Finally, we allow this tree to be traversed, with backtracking, at different creativity (temperature)
levels at different depths to explore alternative novel solutions, related ideas as well as refined versions of the
original problem statement.

We show that LLMs can be successfully tuned to support this traversal with a small number of pre-existing problem-solution mappings,
and that the problem and solution spaces can be effectively explored within a semantic region controlled by LLM temperature.

As a proof of concept we also implement and integrate this exploration approach in a Slack bot to assist innovators in their
ideation process.

The rest of this paper is organized as follows. We first discuss related work in Section~\ref{sec:relatedwork}. We then present the model and algorithm underlying
our approach in Sections~\ref{sec:model} and~\ref{sec:algorithm}. Then we evaluate our
model training and exploration with a public data set in Section~\ref{sec:evaluation}, and describe an
early system prototype in Section~\ref{sec:implementation}. Finally, we provide
concluding remarks in Section~\ref{sec:conclusion}.

\section{Related Work}\label{sec:relatedwork}
Machine-, and AI-, assisted ideation 
and creativity support has been explored extensively in the field
of HCI~\cite{hwang2021,shneiderman2002,yu2011}, 
including application of LLMs to
generate ideas~\cite{di2022,chakrabarty2023,liu2023}.

Creativity support tools have been successfully
used to simply connect creators and allow them to refine
each others' ideas collaboratively~\cite{yu2011}. Early creativity
support tools~\cite{shneiderman2002} targeted refining of existing innovations, 
and exploration of related work to accelerate innovation. 

The impact that these tools have on the ideas produced has been investigated in a 
number of user studies, with interesting design lessons gained.
For example, an AI-bot ideation
collaborator could improve both the quantity and quality
of generated ideas compared to human facilitators, by
removing social pressure, anxiety and bias~\cite{hwang2021};
and the LLM inference times while recursively exploring 
related research problems could pace the ideation
process for a better and more creative ideation 
experience~\cite{liu2023}. 

Using LLMs as a creative writing assistant was explored in
~\cite{chakrabarty2023}, where user studies showed that the
LLM was considered helpful in particular in the writing translation 
(review) task.
Authors provide a genre and a plot and the LLM can produce a draft
and revise it given writer instructions. Study participants found
the tool to be most helpful as an editor as opposed to idea generator.

In~\cite{di2022} the authors propose a tool that leverages prompt 
engineering to expand, rewrite and combine innovations suggested by users,
demonstrating the multi-, and general-purpose NLP capabilities of
LLM foundation models, which we also exploit.

A train-of-thought model for problem solving in LLMs, called Tree-of-Thought (ToT),
was proposed in~\cite{long2023}. ToT was designed as a fully automatic process
to solve math-like problems using logical rules to check which path to take next in
a tree of LLM interactions. Our work exposes this exploration and allows the end-user
not only decide on which path to explore next, but learn from alternative approaches
to tackling the same problem.

None of these works studied the use of internally fine-tuned LLMs for
targeted idea exploration and semantic traversal in custom innovation
problem-to-solution, and solution-to-problem domains.

\section{Model}\label{sec:model}
\subsection{Problem and Solution Statement spaces}
\begin{figure}[htp]
        \centering
                \includegraphics[width=\textwidth]{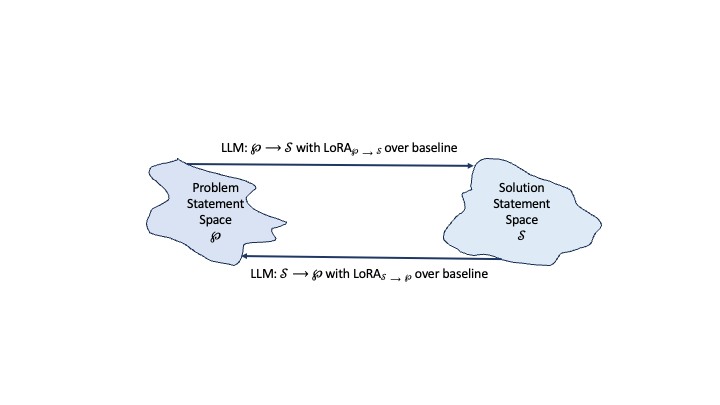}
	\caption{The pair of LLM-based mappings between the spaces of problem statements and solution statements.}
	\label{fig:llmprobsoln}
\end{figure}

We model our processing pipeline in terms of a pair of mappings, one from the space of problem statements to the space of solution statements, and the other from the space of solution statements to the space of problem statements, as shown in Fig.~\ref{fig:llmprobsoln}.  We emphasize that these two mappings are not inverses of one another, as will become clearer from the discussion below.

Before we begin describing the details of our model, we note that there exist mappings from text (whether at the level of characters, $n$-grams, words, or even sentences) to high-dimensional dense real-valued vectors.  For the purposes of the discussion below, the details of how these mappings are defined are not necessary, but we remark in passing that the mappings themselves are \emph{learned} from text using neural networks~\cite{arsanjani2023}.

The mapped vectors are called \emph{embeddings}.  These vectors have dimension ranging from several hundred to a few thousand.  For example, the dimension of the embedding\\\texttt{text-embedding-3-small} used by OpenAI in GPT-3.5 (the engine behind the ChatGPT service) is $1536$.~\footnote{https://platform.openai.com/docs/guides/embeddings} Long texts are split up into shorter text segments, each of which is embedded using such a mapping.

What we loosely call the space of problem statements, denoted $\wp$, is actually a subset\footnote{It is not the entire space because the text vocabulary and grammar mean that not all points in the embedding space correspond to valid text segments or tokens.} of the cartesian product of as many embedding spaces as the maximum number of text segments that a problem statement text is split up into for embedding.  Similarly, what we call the space of solution statements, denoted $\mathcal{S}$, is a subset of the cartesian product of as many embedding spaces as the maximum number of text segments that a solution statement text is split up into for embedding.  Although it is customary to use the same embedding mapping for problem statement text and solution statement text, the maximum number of text segments that problem statement text is split up into may differ from the maximum number of text segments that solution statement text is split up into.  In other words, the dimensions of $\wp$ and $\mathcal{S}$ may not be equal.

\subsection{Mapping between Statement spaces using LLMs}
The mapping from, say, $\wp$ to $\mathcal{S}$ is done using a Large Language Model (LLM) implemented using the so-called \emph{Transformer} architecture~\cite{vaswani2017}. Although a baseline LLM will perform a mapping from one embedding space to another, its output is not tailored to the problem-statement-to-solution-statement mapping use case that is of interest to us here.  Therefore, we adapt the baseline pre-trained LLM to provide outputs better suited to this particular use case by \emph{fine tuning}~\cite{liu2022}, i.e., by further training it on a smaller set of (problem-statement, solution-statement) examples.  

Modifying all parameters of the baseline LLM via further training is computationally expensive.  So we only modify the weights of its Transformer matrix, and this modification takes the computationally lightweight form of a low-rank update added to the Transformer matrix weights.  This technique is called Low-Rank Adaptation (LoRA)~\cite{hu2022}. Another advantage of LoRA is that because it is a low-rank update that is simply added to the Transformer matrix, we can maintain multiple such LoRA low-rank update matrices (each applying to a specific fine-tuning use case), and apply the appropriate LoRA update as needed to the weights of the Transformer matrix in the baseline LLM at the time of inference with the LLM.

For our specific use case, we use a specific baseline LLM and maintain two LoRA updates, one for the mapping from $\wp$ to $\mathcal{S}$, the other for the mapping from $\mathcal{S}$ to $\wp$, as shown in Fig.~\ref{fig:llmprobsoln}.

\subsection{Exploring the Problem Statement spaces}
\label{sec:explore_statement_space}
\begin{figure}[htp]
        \centering
                \includegraphics[width=\textwidth]{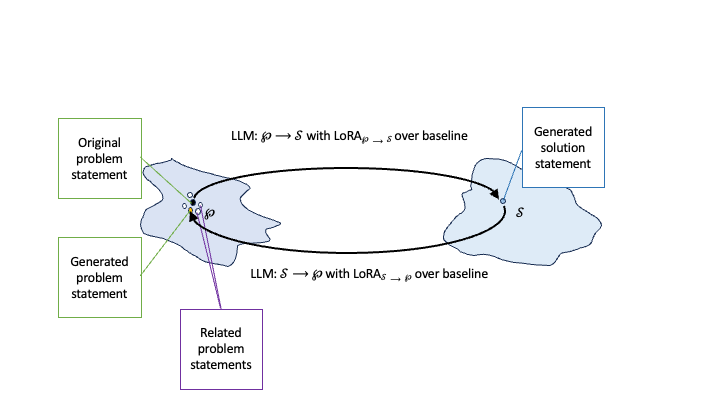}
	\caption{Exploring the Problem Statement space by finding related known problem statements and generating a new problem statement using the two LLM + LoRA mappings of Fig.~ref{fig:llmprobsoln}.}
	\label{fig:llmprobsoln2}
\end{figure}

As already stated earlier, our goal is to develop an ``Ideation Assistant'' that can start with an incomplete or unclear problem statement as input and produce as output related problem statements with the intention of sparking creative ideation in the (human) user interacting with the assistant.  These related problem statements are defined by embeddings corresponding to points in $\wp$ that are ``near'' (by some metric) to the embedding of the original input problem statement.  These points in turn are obtained in two ways, described below and shown in Fig.~\ref{fig:llmprobsoln2}, where problem statement and solution statement embeddings are represented by points in $\wp$ and $\mathcal{S}$ respectively, and semantic similarity is represented by geometric proximity.  For brevity, we refer to these two methods as \emph{selection} and \emph{sampling} respectively.

\subsubsection{Selection} 
Here we \emph{select} problem statements from a knowledge base of known (problem, solution) statement pairs that are semantically related to the input problem statement. We may choose, say, the $4$ known problem statements whose embeddings have the highest cosine similarity to the input problem statement embedding.\footnote{Note that we could alternatively have mapped the input problem statement to a solution statement using the upper LLM and LoRA in Fig.~\ref{fig:llmprobsoln} and searched in $\mathcal{S}$ for known solution statements whose embeddings are near (i.e., have high cosine similarity to) the mapped solution statement.  However, problem statements are usually shorter than solution statements, making their respective semantic distances cheaper to compute, and a single known problem statement may be associated with multiple known solution statements, not all of which may be recovered from a neighbor search in $\mathcal{S}$.  So we perform our search for related statements in $\wp$.} This operation does not require either of the LLM mappings in Fig.~\ref{fig:llmprobsoln} to be applied, as it is entirely performed in $\wp$.

In Fig.~\ref{fig:llmprobsoln2}, these $4$ related known problem statements are represented by the white dots in $\wp$, while the original input problem statement is represented by a black dot and denoted $P_o$.

\subsubsection{Sampling}
Here we \emph{sample} from the probability distribution of the composition of the two LLM + LoRA operations by generating a new, hitherto-unseen problem statement by applying the forward and reverse LLM + LoRA operations in Fig.~\ref{fig:llmprobsoln}.
We begin with the original input problem statement, then use the forward LLM (with corresponding LoRA) of Fig.~\ref{fig:llmprobsoln} to map to a (generated) solution statement, followed by the reverse LLM (with corresponding LoRA) to yield a (generated) problem statement.  The naive expectation may be that this problem statement text will be identical to the original input problem statement text.  However, the randomness of the LLM mapping and the ``creativity'' induced by the temperature parameters (see below) of the two LLMs will make the two problem statement texts (original input, and the generated result of this forward-and-reverse-LLM pass) differ slightly, although they will be semantically very close.  Moreover, the original problem statement could be poorly worded, or even incomplete, with just a phrase and a few keywords, whereas the generated problem statement will be a complete grammatically correct one.  Thus, this operation not only serves to sharpen the human ideator's thinking, it also polishes the original problem statement.

In Fig.~\ref{fig:llmprobsoln2}, the original problem statement, represented by the black dot in $\wp$, is mapped into the generated solution statement represented by the blue dot in $\mathcal{S}$, then mapped back to the generated problem statement represented by the orange dot in $\wp$.  Note that the original and generated problem statements are represented as being semantically very similar but the two embeddings are not identical, meaning that their corresponding texts are not the same.

\subsection{Ensuring proximity of sampling and the role of LLM temperature}
In the above discussion of exploring the two statement spaces by selection and sampling, our implicit assumption is that (semantically) related problem statements (defined as having high cosine similarity between their embeddings) will be mapped by the problem-statement-space-to-solution-statement-space LLM (with corresponding LoRA) to related solution statements, and similarly that (semantically) related solution statements (defined as having high cosine similarity between their embeddings) will be mapped by the solution-statement-space-to-problem-statement-space LLM (with corresponding LoRA) to related problem statements.

\begin{figure}[htp]
        \centering
                \includegraphics[width=3in,height=2in]{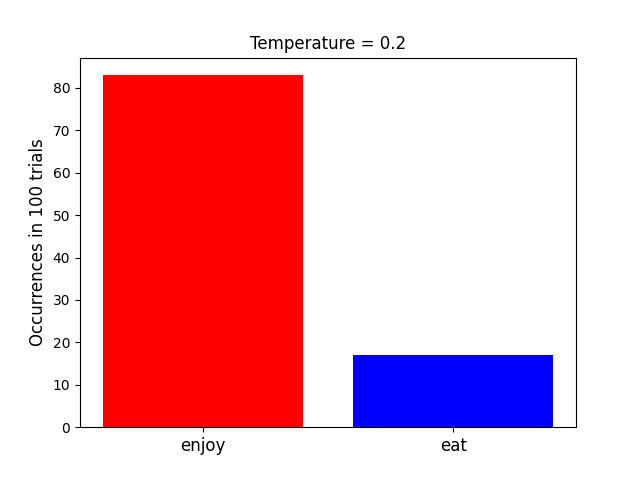}
                \includegraphics[width=3in,height=2in]{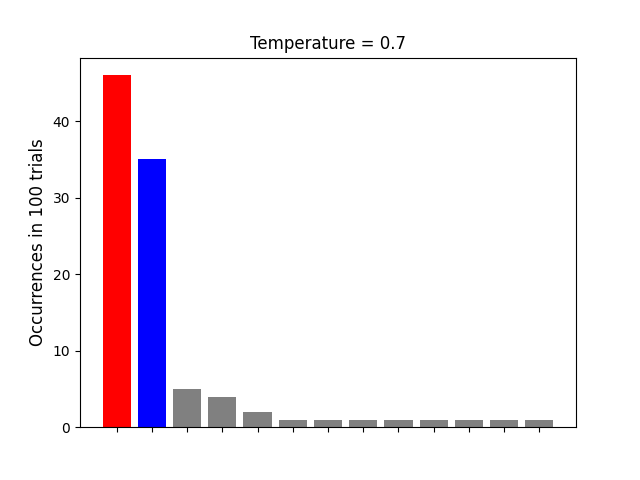}
                \includegraphics[width=3in,height=2in]{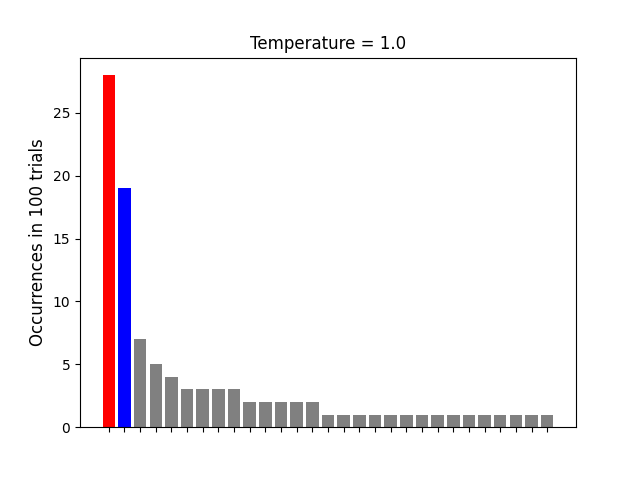}
                \includegraphics[width=3in,height=2in]{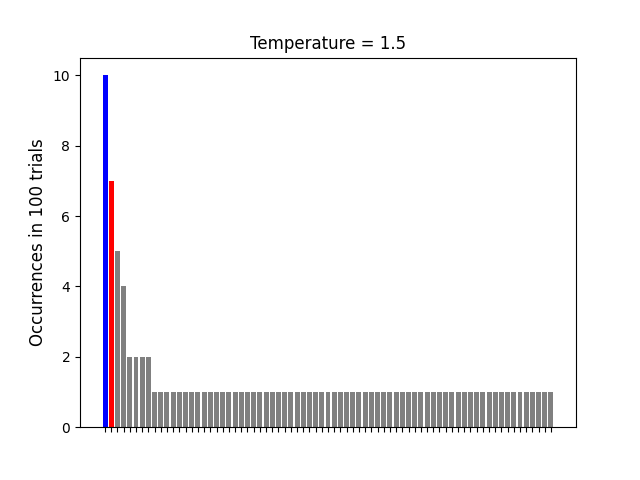}
                \includegraphics[width=3in,height=2in]{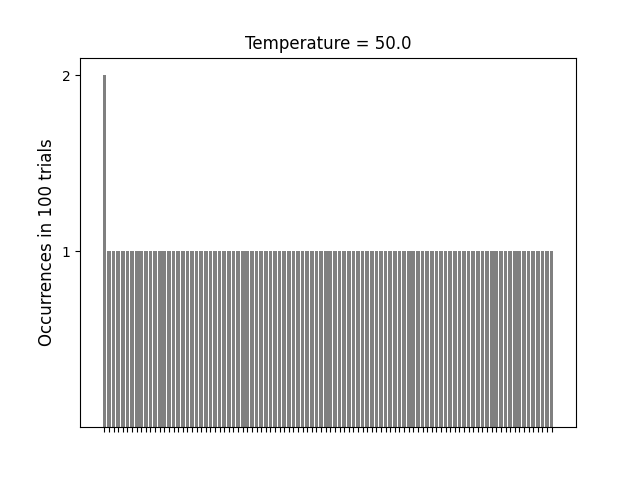}
	\caption{The effect of the temperature parameter on the histogram of the output (single) token of the Pythia-Chat-Base-7B LLM (see Sec.~\ref{sec:pythia-chat-base-7b}) for the next word in the completion prompt ``I like to \textunderscore'' from $100$ different trials.  The two most frequently occurring outcomes, ``enjoy'' and ``eat'' are labeled for the $0.2$ temperature plot but the labels are omitted in subsequent plots for brevity, using instead the red bar for ``enjoy'' and blue for ``eat''.}
	\label{fig:temperature}
\end{figure}

To see why this is plausible, recall that low temperature\footnote{See \url{https://writings.stephenwolfram.com/2023/02/}\\ \url{what-is-chatgpt-doing-and-why-does-it-work/#its-just-adding-one-word-at-a-time} for details.} does not distort the probability distribution over text tokens that is produced by the Transformer inside an LLM during inference and used by the output layer of the LLM to draw the next output token.\footnote{Note that this also means that the LLM is a random mapping.}. In Fig.~\ref{fig:temperature}, we see the effect of changing the temperature parameter on the histogram of the output (single) text tokens generated by the Pythia-Chat-Base-7B LLM we evaluate later in Sec.~\ref{sec:pythia-chat-base-7b}, in response to the prompt to complete ``I like to \textunderscore'', repeated $100$ times.  With a low temperature parameter ($0.2$), the only two outputs are ``enjoy'' and ``eat'', with the former occurring almost $5$ times more frequently than the latter.  As the temperature parameter is increased to $0.7$, then $1.0$, and $1.5$, we see that ``enjoy'' and ``eat'' are still the most frequently-occurring outputs of the LLM, but there are many other outputs also produced and ``enjoy'' and ``eat'' are not the overwhelming number of outcomes out of the $100$ trials anymore.  In fact, for temperature parameter of $1.5$, ``eat'' overtakes ``enjoy'' as the most frequently-occurring outcome.  As the temperature parameter gets larger and larger, the asymptotic effect is to make every outcome of the LLM equally likely.  This is what we notice for the plot with temperature parameter $50$ in Fig.~\ref{fig:temperature}, where there are $99$ distinct observed outcomes out of $100$ trials, all but one occurring exactly once, and ``eat'' and ``enjoy'' happen not to show up at all in these $100$ trials.

There is some empirical evidence to validate the claim that a forward-and-backward pass of the two LLMs (with associated LoRAs) in Fig.~\ref{fig:llmprobsoln} approximately preserves semantic similarity, especially when the \emph{temperature} parameter of the output layer of both LLMs in Fig.~\ref{fig:llmprobsoln} is low.   We notice that when the temperature parameter is set low, a forward-and-backward pass using the two LLMs in Fig.~\ref{fig:llmprobsoln} in succession starting from a given original problem statement yields a problem statement whose text is semantically very similar to the original problem statement, as illustrated in Fig.~\ref{fig:llmprobsoln2}. 

The validity of the above assumption gets increasingly questionable as the temperature parameter increases. However, raising the temperature parameter of an LLM increases the ``creativity'' of its output because the probability distribution over text tokens described above starts to flatten out and be increasingly dispersed, thereby allowing the LLM output layer to draw tokens that would not have been selected at lower values of the temperature parameter.  For a longer discussion on creativity, emergent behavior, and hallucinations in LLMs, see~\cite{huberman2023}.

It follows that, in a use case such as ours where we are looking for assistance in creative ideation, we should try to raise the temperature parameters of both LLMs in Fig.~\ref{fig:llmprobsoln} to the extent possible while still retaining strong relatedness between the input problem statement, the selected related problem statements, and the sampled newly-generated problem statement.  The choice of a suitable range for the temperature parameter for the two LLMs may involve some experimentation and is likely to be different from one problem statement space to another.

\subsection{Wider exploration of the Problem Statement space}
The sequence of steps described in Sec.~\ref{sec:explore_statement_space} yields one newly-generated problem statement for each input problem statement, along with several known problem statements, all of which are related to the input problem statement.  This is useful not only for generating new problem statements, but also as a search tool for known problem statements that are related to the input problem statement.  Note also that we can pair the generated solution statement (obtained from the forward LLM + LoRA mapping applied to the original input problem statement) with the generated problem statement (obtained from the reverse LLM + LoRA mapping applied to the generated solution statement) and archive the generated (problem, solution) statement pair for further exploration if needed.\footnote{For example, the generated (problem, solution) statement pair could be inserted into the enterprise's knowledge base of known (problem, solution) statement pairs after suitable curation by human experts.}

\begin{figure}[htp]
        \centering
                \includegraphics[width=\textwidth]{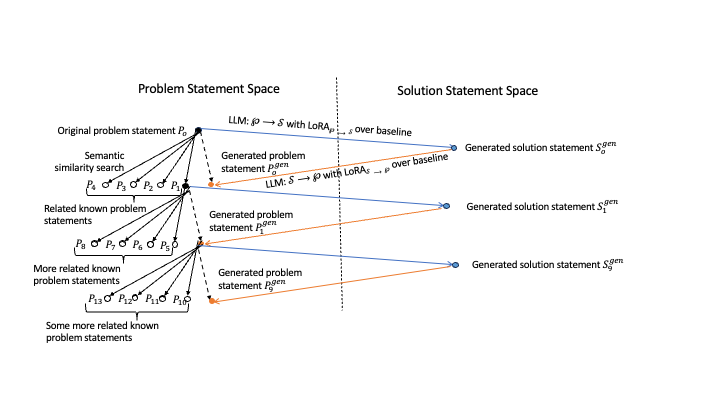}
	\caption{Illustrating a wide depth-first exploration of the Problem Statement space by iteratively applying the two LLM + LoRA mappings of Fig.~\ref{fig:llmprobsoln} to the newest generated problem statement, or alternatively, by searching for known problem statements related to one of the known problem statements discovered in the previous iteration.}
	\label{fig:llmprobsoln3}
\end{figure}

However, what do we do if we want more than one all-new problem statement that is related to the input problem statement?  We could apply the procedure of Sec.~\ref{sec:explore_statement_space} again, this time taking the newly-generated problem statement as the input problem statement.  Alternatively, we could apply the same procedure to any of the known related problem statements discovered by the above step by treating it as the input problem statement.  Note that by applying this procedure twice, we are now selecting or sampling semantic neighbors of semantic neighbors of the original problem statement, so the degree of relatedness of this set of problem statements should be validated carefully. 

The above procedure is illustrated in Fig.~\ref{fig:llmprobsoln3}.  From the original input problem statement $P_o$, we obtain the semantically related known problem statements $P_1, \dots, P_4$, as well as the new generated solution statement $S_o^{gen}$ and the new generated problem statement $P_o^{gen}$.  With reference to Fig.~\ref{fig:llmprobsoln2}, $P_o$ is the black dot, $S_o^{gen}$ the blue dot, and $P_o^{gen}$ the orange dot, and this color scheme is maintained in Fig.~\ref{fig:llmprobsoln3} for consistency.  

As discussed above, at any stage, we could perform the same procedure on either one of the known related problem statements or on the new generated problem statement.  In Fig.~\ref{fig:llmprobsoln3}, we show what happens when we take known related problem statement $P_1$ as the input, yielding related known problem statements $P_5, \dots, P_8$, generated solution statement $S_1^{gen}$, and generated problem statement $P_1^{gen}$.  We also show what happens when, at the next stage, we take the new generated problem statement $P_1^{gen}$ as the input, thereby yielding related problem statements $P_{10}, \dots, P_{13}$, generated solution statement $S_9^{gen}$, and generated problem statement $P_9^{gen}$.  

In short, repeatedly applying the procedure of Sec.~\ref{sec:explore_statement_space} to a chosen problem statement (either known or newly generated) of the immediately prior stage, corresponds to exploring the problem statement space $\wp$ through a depth-first tree-based traversal.  We formalize the procedure illustrated in Fig.~\ref{fig:llmprobsoln3} as an algorithm in the next section.

\section{Algorithm}\label{sec:algorithm}
Taking an operational view of the problem and solution
space traversal we define the following three
operational primitives:
\begin{equation}
\mathsf{sol}(p) \rightarrow s
\end{equation}
\begin{equation}
\mathsf{pro}(s) \rightarrow p
\end{equation}
and
\begin{equation}
\mathsf{rel}(p,k) \rightarrow \{\mathsf{nn}_1(p), \mathsf{nn}_2(p),\dots,\mathsf{nn}_k(p)\} 
\end{equation}
where $p$ is a problem statement, $s$ is a solution generated from a problem
statement, and $\mathsf{nn}_i(p)$ is the $i$th nearest neighbor of the problem statement $p$ (in terms of semantic distance
in embedding space).

In other words, $\mathsf{rel}(p, k)$ represents the $k$ known problem statements (in the idea database) with the highest semantic similarity to (equivalently, least semantic distance from) the original problem statement $p$ in $\wp$. In Fig.~\ref{fig:llmprobsoln2}, these related problem statements are represented by the white dots in $\wp$ near the black dot representing the original problem statement $p$.  In Fig.~\ref{fig:llmprobsoln3}, $\mathsf{rel}(P_o, 4) = \{P_1, \dots, P_4\}$, $\mathsf{rel}(P_1, 4) = \{P_5, \dots, P_8\}$, and $\mathsf{rel}(P_1^{gen}, 4) = \{P_{10}, \dots, P_{13}\}$.

Also, $\mathsf{sol}(\cdot)$ and $\mathsf{pro}(\cdot)$ are respectively the upper and lower LLM + LoRA mappings in Fig.~\ref{fig:llmprobsoln}.  In Fig.~\ref{fig:llmprobsoln3}, $\mathsf{sol}(P_o) = S_o^{gen}$ while $\mathsf{pro}(\mathsf{sol}(P_o)) = P_o^{gen}$. Similarly, $S_1^{gen} = \mathsf{sol}(P_1)$, $P_1^{gen} = \mathsf{pro}(S_1^{gen})$ and $S_9^{gen} = \mathsf{sol}(P_1^{gen})$, $P_9^{gen} = \mathsf{pro}(S_9^{gen})$.

Now, our exploration via the tree traversal depicted in Fig.~\ref{fig:llmprobsoln3} can be formally written as follows:
\begin{equation}
\mathsf{explore}(p,k) \rightarrow \mathsf{sol}(p), \mathsf{pro}(\mathsf{sol}(p)), \mathsf{rel}(p,k),
\end{equation}
where the exploration traversal is defined as:
\begin{enumerate}
\item{$\mathsf{explore}(p,k): s \leftarrow \mathsf{sol}(p), \mathcal{P} \leftarrow \{\mathsf{pro}(\mathsf{sol}(p))\} \cup \mathsf{rel}(p,k)]$}
\item{$\text{Select } p \leftarrow  P_x \text{ for some } x \in \{1,\dots,k+1\}, \quad \text{ where } \mathcal{P} := \{P_1, \dots, P_{k+1}\}$}
\item{Go to step~1}
\end{enumerate}

The collection of $s$ and $p$ values obtained in this process comprise
the solution space and the problem space traversed respectively.
How $x$ is chosen in each iteration is up to the user traversing the space.
Automated traversals could explore the spaces with depth-first or breadth-first
strategies. An example of a Python snippet using random exploration is available 
in Appendix~\ref{python}.

\section{Evaluation}\label{sec:evaluation}
To test and evaluate our approach we use a problem-solution dataset extracted
using the OpenAI API and a public list of company names.

\subsection{Dataset}
No public datasets with problem-solution mappings of innovations are available at a large
scale so some NLP processing is needed. Given the powerful multi-purpose NLP capabilities
of publicly available LLMs such as OpenAI's ChatGPT we opted to leverage LLM prompting
for this step as well. From a list of the top 400 software companies in terms of revenue 
we used the OpenAI API\footnote{with model gpt-3.5-turbo} and the following prompt:
\begin{verbatim}
Provide a short description of the problem the 
company _COMPANY_ solves and how it solves it 
separated by PROBLEM and SOLUTION headers without 
mentioning _COMPANY_ by name.
\end{verbatim}
where we replace the {\it COMPANY} tag with the name of the company for all companies in our list.
We found that we could get the LLM to provide valid solution and problem separations for 313 of the 400 companies tested, and that is thus our evaluation dataset.
Examples of these pairs can be found in Appendix~\ref{company}.

\subsection{Model Training Validation}
\label{sec:pythia-chat-base-7b}
Our end-goal is to produce a bot able to assist with internal innovation and thus interacting with a public API is not an option. Furthermore we
want to use private and confidential prior innovations to train our model. However, given that we hope to extract a fairly complex NLP function of
mapping a problem to a solution (and a solution to a problem) we need a powerful LLM. The solution to this dilemma is {\it fine tuning}
where we take a public foundation model trained on a large public corpus and then fine-tune it with a small set of examples (313 in our case) of
prompt and expected output pairs. As foundation model we opted for the Pythia chat model provided by togethercomputer and
publicly available on Hugging Face\footnote{https://huggingface.co/togethercomputer/Pythia-Chat-Base-7B}. We train both the problem to solution and
the reverse solution to problem pairings with our dataset on 2 Nvidia A40 GPUs with CUDA using the LoRA~\cite{hu2022} approach. Each mapping
is stored in a separate adapter (SolutionProblem vs ProblemSolution) that can be invoked and loaded on demand depending on which transformation to reproduce.
Each direction took about 19 minutes to train with this setup. Both adapters can also be pre-loaded into memory of a single GPU
to make inference efficient.

We also record the problems in a vector database (Redis) which allows for efficient retrieval of nearest-neighbor searches based
on embeddings, again publicly available from Hugging Face\footnote{https://huggingface.co/sentence-transformers/all-mpnet-base-v2}.

Table~\ref{T:sim} shows the average and standard deviation of a test split withheld from the training data and
used in predictions (10 mappings) for pairwise
cosine similarity (ground truth versus generated), and Table~\ref{T:dist} shows the edit distance (Levenshtein) for
the same predictions (generations). 
The following predictors were used in this evaluation:
\begin{itemize}
  \item{{\bf LoRA0.1}. Our fine tuned model with temperature set to $0.1$.} 
  \item{{\bf LoRA1.0}. Our fine tuned model with temperature set to $1.0$.} 
  \item{{\bf Random}. We pick a random problem or solution (depending on what is predicted) from the training data as the prediction.} 
  \item{{\bf OAI}. We feed the same prompt as was used for fine-tuning into the OpenAI API (which was also used to create the dataset).} 
  \item{{\bf OAIT}. We enhance the prompt with a tag such as ``Describe a problem/solution`` to clarify what we want OpenAI to do.} 
\end{itemize}

\begin{table}[htbp]
	\caption{Fine Tuning Similarity. 303/313 train, 10/313 test.}
\begin{center}
\begin{tabular}{|l|c|c|c|c|c|}
\hline
	\textbf{Transformation} &  \multicolumn{5} {c|} {\bfseries Similarity $\mu\pm\sigma$} \\
	 & \textbf{LoRA0.1} & \textbf{LoRA1.0} & \textbf{Random} & \textbf{OAI} & \textbf{OAIT} \\
\hline
	Problem$\rightarrow$Solution & $.85\pm.06$ & $.80\pm.07$ & $.39\pm.12$ & $.77\pm.11$ & $.81\pm.08$ \\
\hline
	Solution$\rightarrow$Problem & $.78\pm.10$ & $.75\pm.12$ & $.23\pm.10$ & $.70\pm.10$ & $.74\pm.14$ \\
\hline
\end{tabular}
\label{T:sim}
\end{center}
\end{table}

\begin{table}[htbp]
	\caption{Fine Tuning Distance. 303/313 train, 10/313 test.}
\begin{center}
\begin{tabular}{|l|c|c|c|c|c|}
\hline
	\textbf{Transformation} &  \multicolumn{5} {c|} {\bfseries Distance $\mu\pm\sigma$} \\
	 & \textbf{LoRA0.1} & \textbf{LoRA1.0} & \textbf{Random} & \textbf{OAI} & \textbf{OAIT} \\
\hline
	Problem$\rightarrow$Solution & $.53\pm.02$ & $.53\pm.03$ & $.55\pm.02$ & $.63\pm.07$ & $.65\pm.08$\\
\hline
	Solution$\rightarrow$Problem & $.49\pm.08$ & $.51\pm.06$ & $.54\pm.02$ & $.73\pm.08$ & $.57\pm.06$ \\
\hline
\end{tabular}
\label{T:dist}
\end{center}
\end{table}

We note that our fine-tuned models are able to replicate the original ground-truth semantically but not
lexically, which is, in fact, desirable in our case of innovation idea exploration, and problem statement
refinement. Furthermore, the more natural problem to solution mapping yields a better result than the reverse
mapping, but both mappings are significantly higher than the random mapping, showing the feasibility of
semantic traversal in both directions. Looking at the OAI results we note that the semantic similarity
is not very high until we tag the prompt with helpful directions on what to output, showcasing that our
fine tuning works. Even with tagged prompts the OAI model, which here was used to produce the
ground truth, and hence should yield near optimal results, is slightly outperformed at the low temperature level $0.1$.
In summary, this experiment showed that our model can maintain semantic fidelity when traversing the solution and 
problem spaces, and that temperature plays a role in the semantic deviation. Next, we will explore more how
traversal is impacted by temperature.

\subsection{Exploration Experiment}
The key aspect of our system that we want to evaluate here is how well it allows us to explore the problem and solution space.
The experiment exploration uses the following procedure to obtain 100 novel generated solutions starting from a single prompt.

\begin{figure}[htp]
        \centering
                \includegraphics[width=\textwidth]{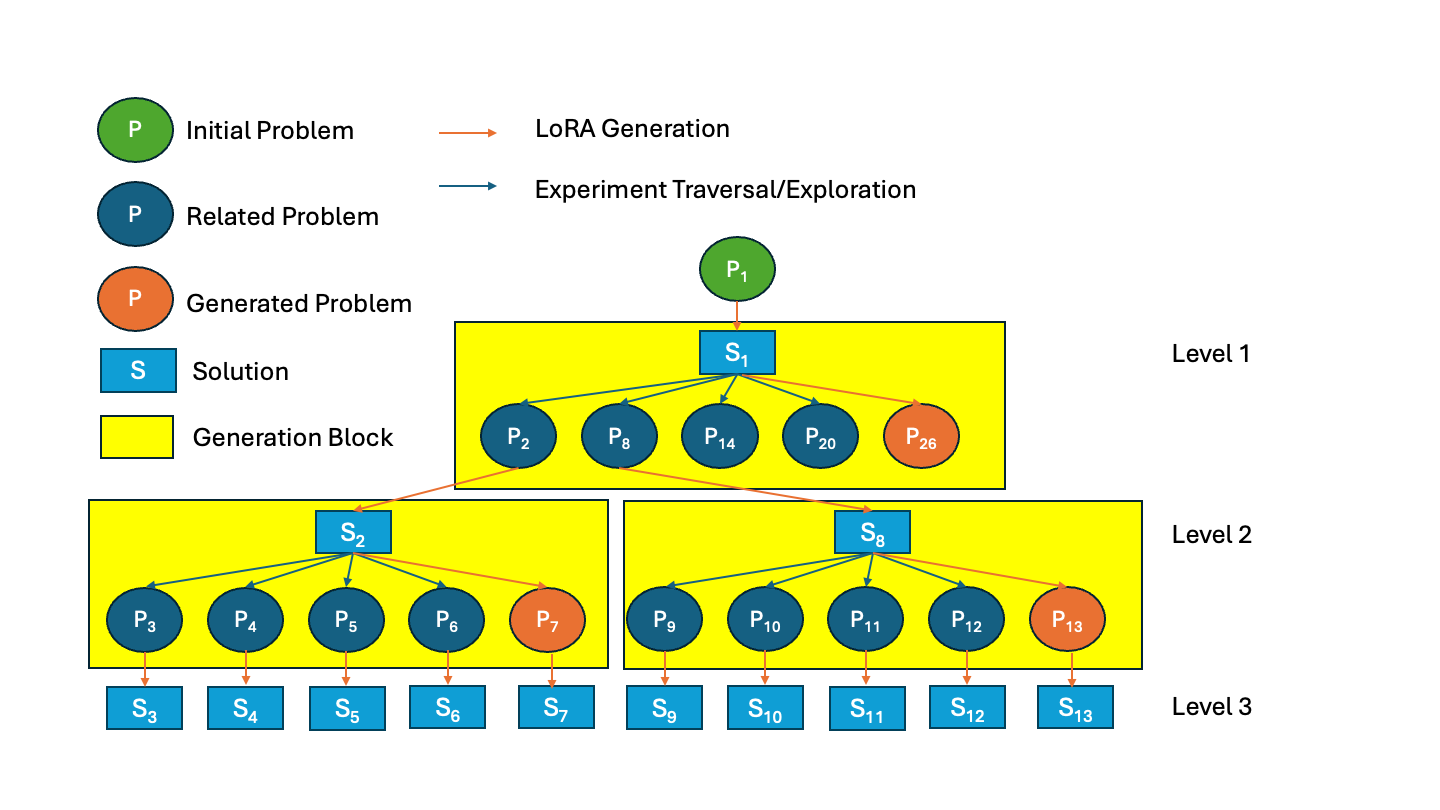}
	\caption{3-level depth-first traversal of problem-solution tree}
	\label{depthtree}
\end{figure}

\begin{enumerate}
	\item{Feed the problem into the prompt of the {\it ProblemSolution} LLM adapter and ask for a solution akin to the format used during LoRA tuning.}
\item{Do a nearest neighbor search of the top 4 related problems in our vectorstore}
\item{Feed the solution generated in step 1 as a prompt into the reverse {\it SolutionProblem} LLM adapter to obtain a novel problem}
\item{Recursively depth-first explore all five problems re-starting with Step 1}
\item{When 100 solutions are generated exit the recursion and compute statistics over the solutions and problems obtained}
\end{enumerate}
Two statistics are computed in the final step for both the list of problems and solutions obtained. 
The average edit distance (Levenshtein) between each unique pairing of solution-solution (or problem-problem)
and the average cosine similarity for the embeddings of each unique pairing of solution-solution (or problem-problem).
Max depth was set to 6. A depth-first search with 3 levels is illustrated in Figure~\ref{depthtree}.

Now we repeat this evaluation while modifying some parameters like the temperature of the LLMs used. Note that
to maintain novelty across many iterations we don't set the temperature to a fixed value but rather pick
a random temperature burst in interval $[0,.1]$ at a given temperature level. 

Examples of solutions generated from problems, and problems generated from solutions with this approach at different temperature levels for the
dataset are shown in Appendix~\ref{solution} and Appendix~\ref{problem} respectively\footnote{the complete dataset is available at https://github.com/cablelabs/llmdata}.

\subsection{Results}
We computed the normalized edit distance and cosine similarity metrics with increasing temperature for
solutions as well as problems for 10 different original problem statements in 10 separate explorations (traversal trees). 
Edit distance 0 and cosine similarity of 1 mean that the two solutions
compared are the same from a text/word perspective, and semantic/embedding perspective respectively. So for our purposes we want the average
edit distance to be as high as possible and the cosine similarity to be as low as possible to obtain more novelty in the solutions.
At the same time we do not want the solutions to drift too far away semantically, so edit distance divergence is more interesting to us.
The metrics are computed as averages across all possible pairs of solutions, and across all possible pairs of problems generated.

\begin{figure}[htp]
        \centering
        \begin{subfigure}[b]{0.49\textwidth}
                \includegraphics[width=\textwidth]{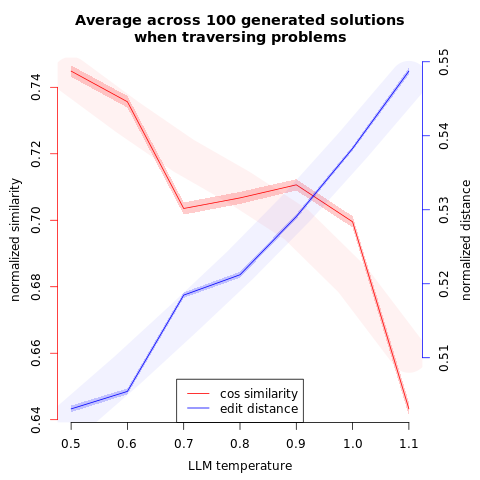}
        \end{subfigure}
        \begin{subfigure}[b]{0.49\textwidth}
                \includegraphics[width=\textwidth]{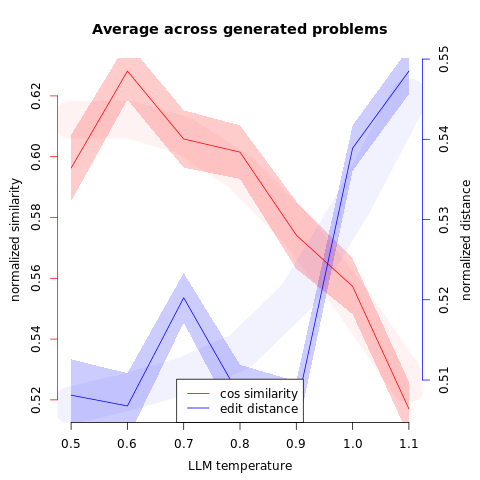}
        \end{subfigure}
	\caption{Solution and Problem Novelty: ``It is difficult to create innovation opportunities without introducing too much process and hampering creativity.''}
	\label{compsolinnovation}
\end{figure}

\begin{figure}[htp]
        \centering
        \begin{subfigure}[b]{0.49\textwidth}
                \includegraphics[width=\textwidth]{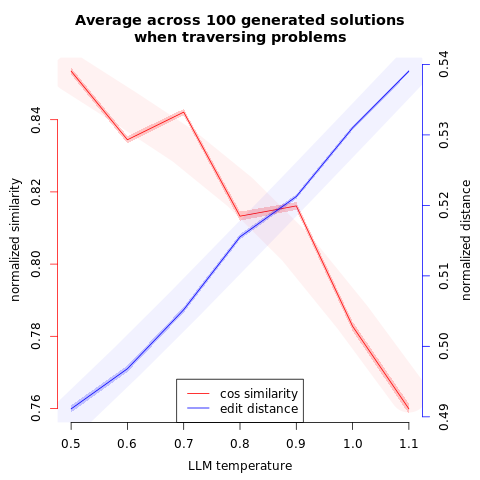}
        \end{subfigure}
        \begin{subfigure}[b]{0.49\textwidth}
                \includegraphics[width=\textwidth]{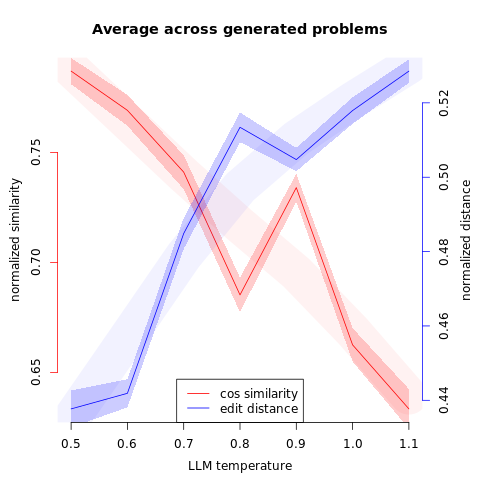}
        \end{subfigure}
	\caption{Solution and Problem Novelty: ``It is difficult to measure employee satisfaction in an unbiased way.``}
	\label{compsolemployee}
\end{figure}

\begin{figure}[htp]
        \centering
        \begin{subfigure}[b]{0.49\textwidth}
                \includegraphics[width=\textwidth]{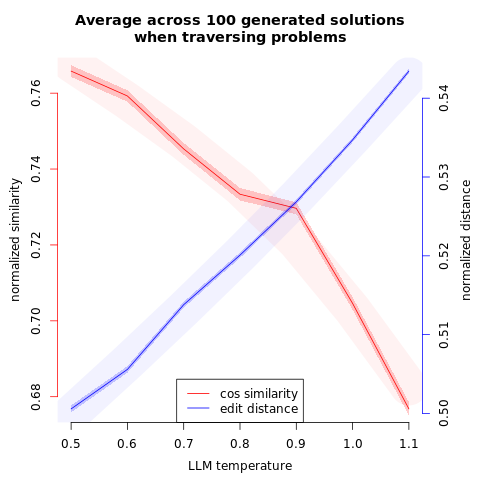}
        \end{subfigure}
        \begin{subfigure}[b]{0.49\textwidth}
                \includegraphics[width=\textwidth]{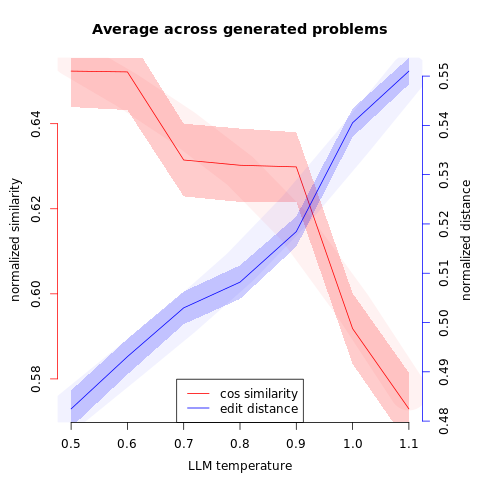}
        \end{subfigure}
	\caption{Solution and Problem Novelty: Average across all original problem statements. See Table~\ref{T:origprob}.}
	\label{compsolcombined}
\end{figure}

We can see in Figure~\ref{compsolinnovation} and Figure~\ref{compsolemployee}, for two example problem statements, 
that solution novelty, defined here lexically as the edit distance\footnote{increasing with temperature} and semantically as the inverse of cosine similarity\footnote{decreasing with temperature}, increases
as expected with temperature.  This trend is even more clear when we compute the averages\footnote{Since different semantic spaces are covered, similarities and distances are still computed within the same problem statement tree, and we then take averages across the statistics computed for each tree.} across all problem statements\footnote{Each exploration at each temperature level involves 100 solution generations, 100 problem generations and 100 nearest neighbor
searches, so this Figure is based on data from a total of $14,000$ LLM generations, and $7,000$ nearest neighbor searches.} in
Figure~\ref{compsolcombined}.

We also observe that for the problem space the same technique works and we get the same novelty
effect on generated problems when increasing the temperature.

\begin{table}[htbp]
  \caption{Original Problem Statements used in Exploration Experiments.}
\begin{center}
\begin{tabular}{|p{.95\linewidth}|}
\hline
Software project timelines are often underestimated, which leads to high costs. \\
\hline
It is difficult to measure employee satisfaction in an unbiased way. \\
\hline
It is not easy for early startups to find a customer base willing to try new technology. \\
\hline
Companies struggle with gaining insights from large volumes and high velocity of data. \\
\hline
It is hard to track and measure customer satisfaction across large geographies. \\
\hline
It is difficult to plan investments in an uncertain economy. \\
\hline
It is difficult to create innovation opportunities without introducing too much process and hampering creativity. \\
\hline
Retaining high-performing talent is hard in competitive emerging markets. \\
\hline
Large machine learning models are expensive and time consuming to train. \\
\hline
Ensuring privacy of customers is difficult while leveraging their data for business insights. \\
\hline
\end{tabular}
\label{T:origprob}
\end{center}
\end{table}

\subsection{Discussion}
We observed that a few original problem statements do not follow the linear novelty-temperature trends as smoothly
as others. This behaviour could be caused by the exploration getting stuck in a semantic region despite
increasing the temperature. In addition to increasing the temperature more drastically these effects could
be mitigated by increasing the $k$ in the top-k nearest neighbor search, or by caching the neighbors already visited
to avoid re-visits. Our approach of not setting a fixed temperature value in the generation call to the LLMs, 
but instead making a uniformly random selection within a set range $\pm.5$, also helps in this aspect,
and hence we carry over this design into the system implementation discussed next.

\section{Implementation}\label{sec:implementation}
\begin{figure}[htp]
     \centering
     \fboxsep=0mm
     \fbox{\includegraphics[width=\textwidth]{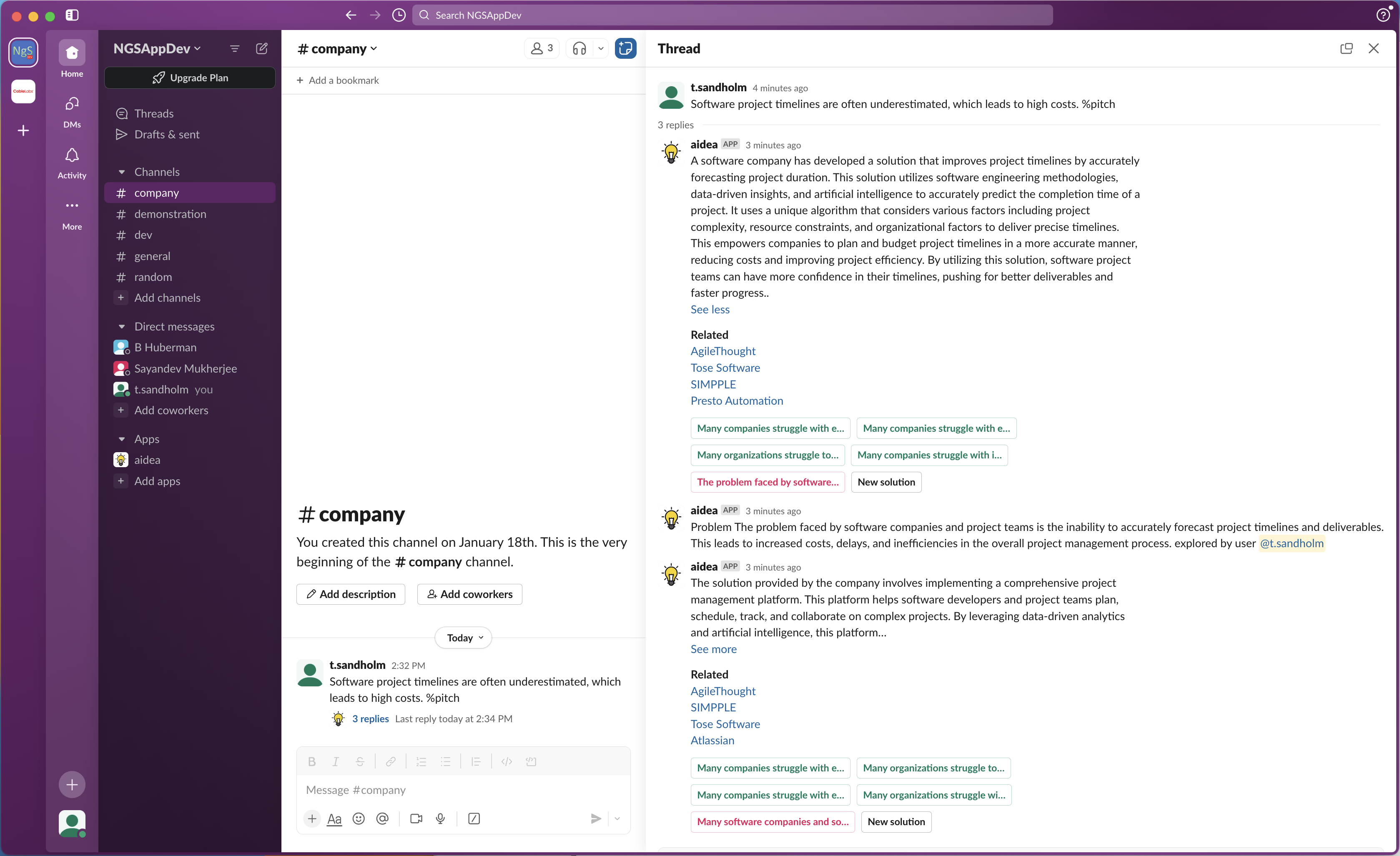}}
     \caption{Slack integration: screenshot of generated problem exploration.}
     \label{ui}
\end{figure}

We have implemented this solution using an integration with Slack where
innovation ideas and related problems to explore can
be invoked with a {\it pitch} tag, see Figure~\ref{ui}.

The four related problems from the semantic search and the reverse
solution to problem mapping are presented as interactive buttons
that any slack user, either the problem submitter or a collaborator
in the same slack workspace can click on to explore the problem.
Given that we use temperature tuned for creativity if you
are not happy with the output for the current problem
explored you can simply ask the bot to generate a new
solution (with new related problems) for the same problem.

The generated problem is colored differently (red) to understand if
existing problem statements are explored or a novel generated one.

The backend uses the node.js Bolt Slack API with web sockets to easily
expose the LLMs to slack from our internal network that has access
to the idea database. The related problems are also presented
as clickable links for exploration.

The LLM implementation uses the OpenChatKit toolkit with LangChain
and a local Redis vectorstore in Docker. To avoid reloading embeddings
and models each time a query comes in they are loaded and exposed to the
Bolt app using a local REST API implemented in Python Flask.

As an experimental feature, we also use the langchain UnstructuredPDF document loader with the RetrievalQA
API (RAG) and FAISS to generate text guiding the innovator how to implement the
solution with different technologies represented as a collection of PDF
documents. We used Wi-Fi and 5G tech reports as an example.
We also use an LLM to summarize and combine or unify the answers from the
different technologies to see if there are network technology 
convergence opportunities.

\begin{figure}[htp]
     \centering
     \includegraphics[width=\textwidth]{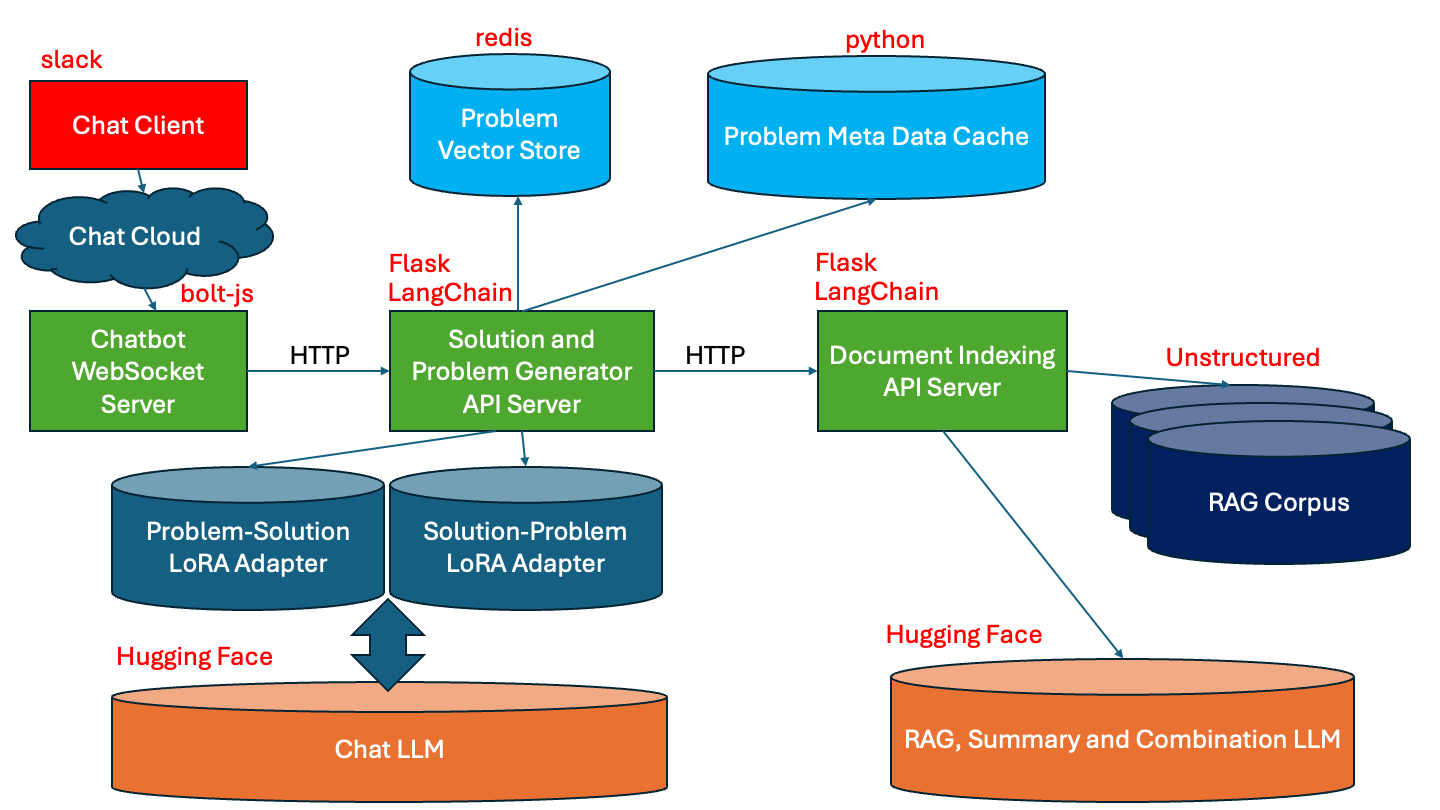}
      \caption{High-level system architecture.}
        \label{architecture}
\end{figure}

\section{Conclusion}\label{sec:conclusion}
We have shown that our idea traversal
approach allows for effective semantically controlled exploration
of a chain-of-thought inspired problem-solution tree.
LLM temperature, LoRA fine-tuning, as well as nearest-neighbor
searches all helped to keep the generated text within a semantic
region while still producing interesting (lexically different), novel and 
creative solutions and problem statements.

\bibliographystyle{IEEEtran}
\bibliography{related}

% Generated by IEEEtran.bst, version: 1.14 (2015/08/26)
\begin{thebibliography}{10}
\providecommand{\url}[1]{#1}
\csname url@samestyle\endcsname
\providecommand{\newblock}{\relax}
\providecommand{\bibinfo}[2]{#2}
\providecommand{\BIBentrySTDinterwordspacing}{\spaceskip=0pt\relax}
\providecommand{\BIBentryALTinterwordstretchfactor}{4}
\providecommand{\BIBentryALTinterwordspacing}{\spaceskip=\fontdimen2\font plus
\BIBentryALTinterwordstretchfactor\fontdimen3\font minus
  \fontdimen4\font\relax}
\providecommand{\BIBforeignlanguage}[2]{{%
\expandafter\ifx\csname l@#1\endcsname\relax
\typeout{** WARNING: IEEEtran.bst: No hyphenation pattern has been}%
\typeout{** loaded for the language `#1'. Using the pattern for}%
\typeout{** the default language instead.}%
\else
\language=\csname l@#1\endcsname
\fi
#2}}
\providecommand{\BIBdecl}{\relax}
\BIBdecl

\bibitem{hwang2021}
A.~H.-C. Hwang and A.~S. Won, ``Ideabot: investigating social facilitation in
  human-machine team creativity,'' in \emph{Proceedings of the 2021 CHI
  Conference on Human Factors in Computing Systems}, 2021, pp. 1--16.

\bibitem{wei2022}
J.~Wei, X.~Wang, D.~Schuurmans, M.~Bosma, F.~Xia, E.~Chi, Q.~V. Le, D.~Zhou
  \emph{et~al.}, ``Chain-of-thought prompting elicits reasoning in large
  language models,'' \emph{Advances in Neural Information Processing Systems},
  vol.~35, pp. 24\,824--24\,837, 2022.

\bibitem{shneiderman2002}
B.~Shneiderman, ``Creativity support tools,'' \emph{Communications of the ACM},
  vol.~45, no.~10, pp. 116--120, 2002.

\bibitem{yu2011}
L.~Yu and J.~V. Nickerson, ``{Cooks or cobblers? Crowd creativity through
  combination},'' in \emph{Proceedings of the SIGCHI conference on human
  factors in computing systems}, 2011, pp. 1393--1402.

\bibitem{di2022}
G.~Di~Fede, D.~Rocchesso, S.~P. Dow, and S.~Andolina, ``{The Idea Machine:
  LLM-based Expansion, Rewriting, Combination, and Suggestion of Ideas},'' in
  \emph{Proceedings of the 14th Conference on Creativity and Cognition}, 2022,
  pp. 623--627.

\bibitem{chakrabarty2023}
T.~Chakrabarty, V.~Padmakumar, F.~Brahman, and S.~Muresan, ``{Creativity
  Support in the Age of Large Language Models: An Empirical Study Involving
  Emerging Writers},'' \emph{arXiv preprint arXiv:2309.12570}, 2023.

\bibitem{liu2023}
Y.~Liu, S.~Chen, H.~Cheng, M.~Yu, X.~Ran, A.~Mo, Y.~Tang, and Y.~Huang, ``{How
  AI Processing Delays Foster Creativity: Exploring Research Question
  Co-Creation with an LLM-based Agent},'' \emph{arXiv preprint
  arXiv:2310.06155}, 2023.

\bibitem{long2023}
J.~Long, ``Large language model guided tree-of-thought,'' \emph{arXiv preprint
  arXiv:2305.08291}, 2023.

\bibitem{arsanjani2023}
\BIBentryALTinterwordspacing
A.~Arsanjani, ``{The Evolution of Text Embeddings},'' 2023. [Online].
  Available:
  \url{https://dr-arsanjani.medium.com/the-evolution-of-text-embeddings-75431139133d}
\BIBentrySTDinterwordspacing

\bibitem{vaswani2017}
A.~Vaswani, N.~Shazeer, N.~Parmar, J.~Uszkoreit, L.~Jones, A.~N. Gomez, L.~u.
  Kaiser, and I.~Polosukhin, ``{Attention is All you Need},'' in \emph{Advances
  in Neural Information Processing Systems}, I.~Guyon, U.~V. Luxburg,
  S.~Bengio, H.~Wallach, R.~Fergus, S.~Vishwanathan, and R.~Garnett, Eds.,
  vol.~30.\hskip 1em plus 0.5em minus 0.4em\relax Curran Associates, Inc.,
  2017.

\bibitem{liu2022}
H.~Liu, D.~Tam, M.~Mohammed, J.~Mohta, T.~Huang, M.~Bansal, and C.~Raffel,
  ``Few-shot parameter-efficient fine-tuning is better and cheaper than
  in-context learning,'' in \emph{Advances in Neural Information Processing
  Systems}, A.~H. Oh, A.~Agarwal, D.~Belgrave, and K.~Cho, Eds., 2022.

\bibitem{hu2022}
E.~J. Hu, Y.~Shen, P.~Wallis, Z.~Allen-Zhu, Y.~Li, S.~Wang, L.~Wang, and
  W.~Chen, ``{Lo{RA}: Low-Rank Adaptation of Large Language Models},'' in
  \emph{International Conference on Learning Representations}, 2022.

\bibitem{huberman2023}
B.~A. Huberman and S.~Mukherjee, ``{Hallucinations and Emergence in Large
  Language Models},'' \emph{SSRN preprint SSRN:4676180}, September 2023.

\end{thebibliography}
\newpage
\appendix
\section{Company Dataset LLM Prompt and Output Examples}\label{company}
\begin{table}[htbp]
        %\caption{Prompts and Responses from OpenAI API.}
\tiny
\begin{center}
\begin{tabular}{|p{.99\linewidth}|}
\hline
Provide a short description of the problem the company {\bf Apple} solves and how it solves it separated by PROBLEM and SOLUTION headers without mentioning {\bf Apple} by name.\\
{\bf PROBLEM:} \\
The problem this company tackles is the need for accessible and user-friendly technology. Many individuals struggle with complicated devices and find it challenging to navigate through various applications and features. Additionally, there is a demand for seamless integration between different devices and systems.\\
{\bf SOLUTION:} \\
The company offers a range of products and services that provide a simplified and intuitive user experience. Their devices are designed with a user-friendly interface, making it easier for individuals to interact with technology. Furthermore, their ecosystem allows for seamless integration among different devices, providing a cohesive and streamlined experience across multiple platforms. Through their innovative solutions, the company aims to bridge the gap between technology and its users, ensuring that anyone can access and utilize their products effortlessly. \\ 
\hline
Provide a short description of the problem the company {\bf Alphabet (Google)} solves and how it solves it separated by PROBLEM and SOLUTION headers without mentioning {\bf Alphabet (Google)} by name. \\
{\bf PROBLEM:} \\
Individuals and businesses often struggle with finding and organizing information from a vast amount of sources on the internet. The complex and ever-expanding nature of the online world poses a challenge in navigating and accessing relevant data efficiently. This can lead to information overload, wasted time, and difficulty in decision-making.\\
{\bf SOLUTION:} \\
The company has developed advanced algorithms and technologies to index, analyze, and present information from the internet in a fast and accurate manner. Through its search engine, the company enables users to effortlessly search for and retrieve relevant information from billions of web pages. By employing various ranking factors and machine learning techniques, the solution provides users with the most helpful and contextually accurate search results. Additionally, the company offers various additional services and tools to further assist individuals and businesses in their information management, productivity, and communication needs.\\ 
\hline
Provide a short description of the problem the company {\bf Microsoft} solves and how it solves it separated by PROBLEM and SOLUTION headers without mentioning {\bf Microsoft} by name. \\
{\bf PROBLEM:} \\
Many people and businesses struggle with using and managing their computer systems efficiently. They often face challenges such as disorganization, security vulnerabilities, lack of communication, and compatibility issues.\\
{\bf SOLUTION:} \\
To address these problems, a technology company offers a wide range of software products and services that aim to streamline computer operations, enhance productivity, foster collaboration, and secure information. Their solutions include operating systems, productivity tools, communications platforms, and cloud services. By developing and delivering innovative software solutions, the company helps users and organizations address their computer-related challenges and improve their overall efficiency and effectiveness.\\
\hline
Provide a short description of the problem the company {\bf IBM} solves and how it solves it separated by PROBLEM and SOLUTION headers without mentioning {\bf IBM} by name. \\
{\bf PROBLEM:} \\
Many organizations struggle with effectively managing and analyzing vast amounts of data generated in today's digital world. Traditional methods of data handling and processing often prove inefficient, leading to missed opportunities for growth, insights, and cost savings. Additionally, the complexity and variety of data sources further compound these challenges and hinder decision-making processes.\\
{\bf SOLUTION:} \\
To address this problem, a company offers a comprehensive data management and analytics solution. Their solution leverages advanced technologies, such as artificial intelligence and machine learning, to streamline the organization's data operations. The platform integrates a wide range of data sources, providing a centralized and unified view of the organization's data assets. By employing advanced analytics algorithms, the solution helps businesses uncover valuable insights and patterns within their data. These insights empower organizations to make data-driven decisions and establish a competitive edge in their industry. Furthermore, the solution enables efficient data governance, privacy, and security measures to ensure compliance with regulations and protect sensitive information. Overall, the company's solution enables organizations to harness the power of their data, optimize operations, improve customer experiences, and drive innovation across various sectors. \\
\hline
Provide a short description of the problem the company {\bf Oracle} solves and how it solves it separated by PROBLEM and SOLUTION headers without mentioning {\bf Oracle} by name. \\
{\bf PROBLEM:} \\
Businesses generate a vast amount of data ranging from customer information to financial records. However, managing and organizing this data can be a challenging task. Additionally, as businesses grow and expand, the task of maintaining and integrating multiple databases from different departments becomes increasingly complex. This can lead to disjointed information, inefficiencies, and difficulties in making informed decisions in a timely manner.\\
{\bf SOLUTION:} \\
To overcome these challenges, a comprehensive database management system is required. Such a system enables businesses to efficiently store, organize, and retrieve data, while also ensuring its security and integrity. It provides a centralized platform that integrates various databases, allowing for seamless sharing of information across departments. Moreover, advanced analytics and reporting capabilities empower businesses to derive meaningful insights from their data, aiding in better decision-making processes. By offering a holistic solution for database management and analysis, this technology helps businesses streamline their operations, improve efficiency, and drive growth. \\
\hline
Provide a short description of the problem the company {\bf Salesforce} solves and how it solves it separated by PROBLEM and SOLUTION headers without mentioning {\bf Salesforce} by name. \\
{\bf PROBLEM:} \\
Companies face difficulties in managing their customer data, leads, and sales processes efficiently. Sales teams often struggle with disorganized information, lack of collaboration, and difficulty in tracking progress and pipelines. This can lead to missed opportunities, poor customer experiences, and a decrease in overall sales effectiveness.\\
{\bf SOLUTION:} \\
A company provides a comprehensive software solution that solves these challenges by offering a centralized platform for managing customer data, leads, and sales processes. This solution allows companies to track and analyze their sales pipelines, collaborate with team members, automate repetitive tasks, and streamline the overall sales process. With easy access to real-time data and personalized insights, sales teams can make informed decisions, enhance customer interactions, and increase their overall productivity. This software solution offers scalability and customization options, catering to the unique needs of every business, and ultimately improving their sales efficiency and driving growth.\\
\hline
Provide a short description of the problem the company {\bf Automatic Data Processing} solves and how it solves it separated by PROBLEM and SOLUTION headers without mentioning {\bf Automatic Data Processing} by name. \\
{\bf PROBLEM:} \\
Companies face challenges in managing and processing their employee data, payroll, and other human resources tasks efficiently. Traditional manual methods can result in errors, delays, and compliance issues.\\
{\bf SOLUTION:} \\
The company provides a comprehensive and automated platform that streamlines and simplifies employee management and human resources processes. Their solution integrates payroll, benefits administration, time and attendance tracking, and other HR functions into a single system. This eliminates manual data entry, reduces errors, ensures compliance with regulations, and enhances overall efficiency.\\
\hline
%Provide a short description of the problem the company {\bf Intuit} solves and how it solves it separated by PROBLEM and SOLUTION headers without mentioning {\bf Intuit} by name. \\
%{\bf PROBLEM:} \\
%Many individuals and small businesses struggle with managing their finances, tax obligations, and bookkeeping. Keeping track of income, expenses, and ensuring compliance with complex tax regulations can be overwhelming for many people, leading to increased stress and potential financial mistakes. \\
%{\bf SOLUTION:} \\
%Our company provides a comprehensive suite of financial management software and services that simplify and automate these tasks. Through our intuitive and user-friendly platforms, individuals and businesses can easily track and categorize their income and expenses, generate financial reports, and seamlessly navigate the intricate tax filing process. Our software also offers features to integrate with various accounting systems, making bookkeeping more efficient and accurate. By streamlining these financial processes, we empower our users to better manage their finances, save time, reduce errors, and ultimately achieve their financial goals.\\
%\hline
\end{tabular}
\end{center}
\end{table}

\newpage
\section{Problem to Solution Mapping Examples}\label{solution}
\begin{table}[htbp]
\tiny
\begin{center}
\begin{tabular}{|l|p{.45\linewidth}|p{.45\linewidth}|}
\hline
{\bf Temp} & {\bf Problem} & {\bf Generated Solution} \\
\hline
0.5 & The problem is that companies need to effectively manage and optimize their product development processes, including complex engineering tasks, collaboration across teams, and efficient deployment of resources. Traditional methods often result in inefficiencies, data fragmentation, and lack of visibility, hampering productivity and innovation. &   By leveraging advanced technologies, such as cloud-based platforms and artificial intelligence (AI), the company offers a comprehensive solution to address the problem of product development management. Their solution offers a suite of tools and services that streamline collaboration, automate workflows, and enable real-time monitoring and analysis of project data. By integrating various components like design tools, simulation, and engineering simulation, the solution enables companies to efficiently manage the entire product development lifecycle. Additionally, the platform provides a centralized hub for project stakeholders to track and manage tasks, assign responsibilities, and track progress. This enables a more coordinated, efficient, and data-driven approach to product development. \\
\hline
0.6 & Many companies struggle with effectively managing their software development processes. They often face challenges in coordinating tasks, tracking progress, and ensuring efficient collaboration among team members. This can lead to delays, miscommunication, and reduced productivity. &   By providing a comprehensive software development platform, this company enables businesses to effectively manage their software development processes. Their solution includes robust collaboration tools, task management systems, and integrated development environments. These tools streamline communication, promote collaboration, ensure timely completion of tasks, and drive overall efficiency in software development. By providing a central hub for team members to access and update information, the solution helps companies tackle the challenges of coordinating multiple tasks, tracking progress, and ensuring smooth collaboration. With its integrated development environments, businesses can easily prototype, code, and test their software, ultimately leading to better quality and faster time to market. \\
\hline
0.7 & Many companies struggle with effectively managing their software development projects. They often face challenges in terms of project scope, requirements gathering, development timeline, and coordination amongst team members. Due to these complexities, projects frequently experience delays, budget overruns, and poor product quality. &   The company offers a comprehensive software development platform that helps organizations streamline and optimize their software development processes. By providing robust tools, the platform assists teams in effectively managing project scope, requirements gathering, development timeline, and overall team coordination. This helps companies meet deadlines, minimize budget overruns, and deliver high-quality products with reduced risk and overall efficiency. With its comprehensive suite of tools and expertise, the company empowers organizations to developer innovative solutions efficiently and effectively. \\
\hline
0.8 & Many companies struggle with managing projects, tasks, and team collaboration efficiently. Traditional methods of tracking progress, communicating updates, and coordinating resources can be cumbersome and lead to confusion and delays. This lack of streamlined workflow can result in decreased productivity and hinder the overall success of a project. &   The company provides a comprehensive project management platform that integrates various tools and features to replace the traditional method of managing projects. Their platform enables team collaboration through real-time communication, task management, and project boards. It ensures accurate tracking of tasks, organizes critical documentation, and provides updates to stakeholders in a centralized location. With this solution, companies can streamline project management, reduce confusion, enhance collaboration, and achieve better efficiency. By reducing time wasted on administrative tasks and improving communication, this platform enhances productivity, reduces errors, and ultimately drives project success. 
The company's solution empowers teams to effectively manage their projects, stay informed, and make informed decisions. \\
\hline
0.9 & Many organizations struggle to effectively manage their projects and track progress. They find it challenging to collaborate and communicate efficiently, which leads to delayed deadlines, miscommunication, and ultimately hampers productivity. Additionally, accessing relevant project information and ensuring alignment across teams can be a cumbersome task. &   The company provides an integrated project management system that aids organizations in streamlining tasks, ensuring progress visibility, and fostering collaboration. Their software enables teams to efficiently mobilize resources, track progress, and monitor deliverables in real-time. It also allows for effective communication, allowing for real-time updates, comments, and reviews, ensuring clarity and transparency throughout the project lifecycle. By offering a user-friendly and comprehensive platform, the company enhances productivity and enables organizations to achieve their goals faster. \\
\hline
1.0 & Many organizations struggle with effective team collaboration and project management due to a lack of proper communication channels and disjointed workflows. Teams may find it difficult to share information, track progress, and coordinate efforts, leading to delays, misalignment, and inefficiencies. &   By leveraging innovative technology, a service-oriented company offers a comprehensive solution for effective team collaboration and project management. Their platform provides seamless communication channels, facilitating timely and secure chatting, voicing, and sending of various project-related documents. Additionally, their platform provides powerful tools such as project backlog interfaces, Kanipboxes, and join.me for visually organizing project teams and facilitating collaboration. Furthermore, their services aim to streamline workflows by automating and organizing project tasks, leading to higher efficiency and productivity. With this solution, organizations can ensure that their teams work cohesively, efficiently, and on the right track, thereby fostering greater success in achieving their goals. \\
\hline
1.1 & Many companies struggle with managing projects, tasks, and team collaboration efficiently. Traditional methods of tracking progress, communicating updates, and coordinating resources can be cumbersome and lead to confusion and delays. This lack of streamlined workflow can result in decreased productivity and hinder the overall success of a project. &   To address this problem, a company has developed a breakthrough software solution that enables efficient project management. This solution enables individuals and teams to organize projects, assign tasks, track progress, and share documents effortlessly. With its intuitive interface and robust features, this software automates and simplifies project management, significantly reducing friction and enhancing collaboration. It presents a clear overview of resources, upcoming deadlines, and progress updates, facilitating informed decision-making and smooth execution. With this solution, companies can confidently deliver better project outcomes while enhancing productivity and efficiencies. \\
\hline

\end{tabular}
\end{center}
\end{table}

\newpage
\section{Solution to Problem Mapping Examples}\label{problem}
\begin{table}[htbp]
\tiny
\begin{center}
\begin{tabular}{|l|p{.45\linewidth}|p{.45\linewidth}|}
\hline
{\bf Temp} & {\bf Solution} & {\bf Generated Problem} \\
\hline
0.5 &   The company provides a comprehensive project management solution that integrates various tools and features to tackle the challenges associated with managing multiple projects simultaneously. Their platform enables teams to create and track tasks, track progress, assign roles and permissions, create milestone check-ins, and communicate updates efficiently. This solution simplifies project management, enhances collaboration, and fosters efficient workflow, leading to increased productivity and success. By streamlining project management, the company empowers teams to effectively manage their projects and achieve their desired outcomes. & Many companies struggle with managing multiple projects and tasks simultaneously. They often face difficulties in tracking and organizing tasks, making sure everyone is aware of their respective responsibilities, and ensuring timely communication and collaboration across teams and departments. \\
\hline
0.6 &   The company provides a comprehensive project management solution that helps organizations manage their projects effectively. By facilitating collaboration, tracking progress, and providing visibility into the overall project status, they enable teams to work together efficiently, stay on track, and ensure timely completion. Their solution also offers features such as task management, time tracking, project milestones, and reporting tools, making it easy for teams to track their progress, monitor deadlines, and make necessary adjustments. Furthermore, their platform integrates with other tools and systems, further streamlining communication and ensuring alignment across teams. Overall, this solution helps organizations optimize project management, improve productivity, and achieve desired outcomes. & Many organizations struggle with managing their project management and collaboration efforts effectively. They often face challenges in coordinating teams, tracking progress, and ensuring timely project completion. \\
\hline
0.7 &   A company offers a comprehensive cloud-based collaboration and project management platform that helps organizations streamline communication, enhance team productivity, and overcome disparate collaboration challenges. Their platform provides a centralized hub for organizing and organizing team communication, ensuring seamless collaboration across devices and platforms. It offers features such as online file sharing, project management tools, task management systems, and real-time chat channels, all integrated into a unified platform. This solution not only improves collaboration and productivity but also ensures that projects are executed effectively and efficiently. By offering a seamless, user-friendly interface, this company empowers teams to work together efficiently and effectively. & Many companies struggle with ineffective collaboration and productivity issues. They often face challenges in organizing and maintaining communication across different teams and platforms, resulting in disjointed workflows and a lack of efficiency. \\
\hline
0.8 &   The company provides a solution that addresses these challenges by leveraging cloud-based tools and communication platforms. By providing a centralized platform, companies can empower their teams with tools for organizing project tasks, setting priorities, tracking progress, and communicating updates effectively. This centralized system eliminates the need for multiple tools or systems, reducing the risk of data loss or miscommunication, and enabling smooth project coordination and overall success. Additionally, the software facilitates real-time collaboration with team members, allowing for seamless communication across different locations and time zones. Overall, the solution helps companies optimize project management, improving productivity and achieving success in their overall goal. & Many companies struggle with organizing and managing their project-based work and communication efficiently. They face challenges such as difficulties in tracking tasks, managing priorities, and ensuring timely updates across teams or locations. \\
\hline
0.9 &   The company offers a comprehensive platform that facilitates team collaboration and project management. Their software enables streamlined communication, real-time file sharing, and automatic project tracking. With this solution, organizations can improve productivity and efficiency, allowing teams to focus on their mission without having to worry about disjointed tasks. Additionally, the platform provides constantly updated data and insights, ensuring transparency and understanding of tasks progress. This solution empowers and supports efficient and responsive project management, leading to better outcomes for organizations and their stakeholders. & Many companies and organizations struggle with inefficient communication, disjointed task management, and lack of real-time collaboration. This creates difficulties in team coordination, project management, and accurate tracking of tasks. \\
\hline
1.0 &   The company's solution offers a comprehensive software solution designed to streamline and automate the product development cycle. It includes tools for engineering design and metrology analysis, as well as collaboration tools for collaboration across teams and inside organizations. Through these tools, companies can effectively manage engineering tasks, such as toolpath planning, nesting, six-sigma validation, surface modeling, fabrication, and pre-production inspection. Additionally, the solution automates the deployment of resources by allowing for faster iteration cycles and precise planning, minimizing the risk of costly errors and optimizemobility. With its robust solution, companies can improve productivity, enhance innovation, and deliver better products faster. & Companies involved in the production of physical goods face numerous challenges in delivering high-quality products on-time and within budget. One of the key obstacles is the lack of efficient and streamlined methods for delivering product designs into reality. \\
\hline
1.1 &   Our company offers a comprehensive software development platform that streamlines the whole process. Our platform provides scalable agile development tools and features, such as project management and Bug Tracking Systems. By integrating existing systems and pull requests, it enables team members to collaborate effectively and efficiently create comprehensive and verifiable requirements.With our solution, companies can easily capture their requirements, identify common concerns, prioritize features, and define timelines. Through continuous deployment, we automate the production and rollout of new features in a continuous delivery cycle. Our platform integrates with various software infrastructure tools to facilitate collaboration with internal and external resources. & companies often struggle with efficiently capturing and incorporating their software development requirements in their processes and systems. This often leads to a disjointed approach and a lack of clarity on priorities. \\
\hline

\end{tabular}
\end{center}
\end{table}

\newpage
\section{Example Random Exploration with Python API}\label{python}

\begin{lstlisting}[language=Python]
from aidea import Aidea
import random

ai = Aidea()
problem = "Software project timelines are often underestimated, which leads to high costs."
n = 10

for i in range(0,n):
  res = ai.explore(problem)
  solution = res["solution"]
  print(f"{problem}\t{solution}")
  problem = random.choice(res["problems"] + [res["generated_problem"]])
\end{lstlisting}

\end{document}